\documentclass[preprint,12pt, authoryear]{mn2e}
\usepackage{graphicx}
\usepackage{amssymb}
\usepackage{footnote}
\usepackage{threeparttable}
\usepackage{epsfig}
\usepackage{hyperref}
\hypersetup{colorlinks=true,urlcolor=magenta,filecolor=magenta,linktoc=page,citecolor=blue,linkcolor=blue}
\usepackage[usenames,dvipsnames,svgnames]{xcolor}
\usepackage{pslatex}
\usepackage{amsmath}
\usepackage{epstopdf}
\usepackage{multirow}
\usepackage{array}

\title[Novae in Quiescence Phase]{Photoionization modelling of quiescence phase spectra of novae \& symbiotic star}
\author[Mondal et al.]{Anindita Mondal$^{1}$\thanks{E-mail: aanyndeta@gmail.com}, Ramkrishna Das$^{1}$, G. C. Anupama$^{2}$, Soumen Mondal$^{1}$\\
$^{1}$S N Bose National Centre for Basic Sciences, Salt Lake, Kolkata 700 106, India\\
$^{2}$Indian Institute of Astrophysics, II Block Kormangala, Bangalore 560 034, India}
\date{November 2019}

\pagerange{\pageref{firstpage}--\pageref{lastpage}} \pubyear{2019}

\begin{document}

\label{firstpage}

\maketitle
\begin{abstract}
We present results of study, using observed and published spectra in optical region, of few novae (T CrB, GK Per, RS Oph, V3890 Sgr and V745 Sco) in their quiescence phase and a symbiotic star (BX Mon). Observations were made using the facilities available at $2m$ Himalayan Chandra Telescope (HCT). Generally, the spectra show prominent low ionization emission features of hydrogen, helium, iron and oxygen and TiO absorption features due to the cool secondary component; T CrB and GK Per show higher ionization lines. We used photoionization code CLOUDY to model these spectra. From the best-fit models, we have estimated the physical parameters, e.g., temperature, luminosity \& hydrogen density; estimated elemental abundances and other parameters related to the system. By matching the spectra of various giants with the absorption features and from the best-fit, we determined the type of secondaries and also their contribution to the spectra.
\end{abstract}

\section{Introduction}        
Novae are interacting binary systems consisting of a primary white dwarf (WD) and main sequence K-M type secondary. Hydrogen-rich matter from the secondary is accreted on the WD surface as a result of mass transfer, via an accretion disk. When a critical temperature and pressure are reached, hydrogen burning sets in, which soon builds up to a thermonuclear runaway (TNR) reaction that releases a huge amount of energy (in the range $10^{38} - 10^{40} ergs$) in a very short period of time (e.g.  Starrfield et al.  2008;  Yaron et al.  2005), which is commonly known as nova explosion. The explosion causes a sudden rise in optical brightness, generally with an amplitude of $\sim$ 7 to 15 magnitudes in 1-2 days. This is followed by a gradual decline in the light curve and the system goes into the quiescence phase. During the quiescence phase, the system goes back to its previous condition and again the accretion process sets in. Theoretically all novae should undergo repeated outbursts, however, the time between two consecutive outbursts may vary depending on different parameters, e.g., accretion rate, white dwarf mass, type of the companion, etc. Novae that undergo more than one explosion in human life timescales are called recurrent novae (RNe)  and those which show only one outburst in a human-life, are classified as classical novae (CNe). In the case and where the companion is a giant, it is termed as symbiotic stars (Shore et al. 2011).\\
\\
During the outburst phase, the spectra show strong emission lines of hydrogen and other elements. Higher ionization lines of  several elements e.g. helium, oxygen, nitrogen, carbon, iron, nickel, calcium, silicon, argon etc. are seen in the nebular and coronal phase (Bode \& Evans 2008). 
On the other hand, quiescence phase spectra are dominated by the absorption features of the secondary along with 
the emission lines originating from the cooler, outer part of the disk (Williams 1980), also known as chromospheric emission. 
The thermal continuum is produced by the hot WD combined with the inner and hotter portion of the disk.
The emission lines are generated by photoionization mechanism as the gas in the disk is ionized by the energetic photons emitted from the central continuum source and heated up by the residual kinetic energy of the photoelectrons (Ferland 2003). 
Thus the spectral characteristics during the quiescence phase reveal the nature of the secondary; the accretion disk; WD; elemental abundances; type of the object etc.
Thus, such studies are crucial to understanding many aspects of the nova outburst and its evolution. \\
\\
However, in comparison to outbursting novae, quiescent novae have not been studied in adequate details due to several causes.
First, the identification of the quiescent nova is often not unambiguous as novae in the quiescent state are very faint. 
Sometimes, based on their blue colours and H$\alpha$ brightness, they can be identified (Anupama \& Kamath 2012). 
Second. as the systems are faint in quiescence phase, magnitude limits of telescopes restrict us to observe them. 
Third, because of their low brightness, long hours of observation are needed and sometimes signal to noise ratio is low which affects the quality of the spectra. 
In view of this, we have monitored a handful of novae and symbiotic stars which show novae-like variability in quiescence phase.
In this paper, we present the results of our study of the novae T CrB,  GK Per, RS Oph, V3890 Sgr, V745 Sco and the symbiotic star BX Mon as the behaviour of the system is similar to that of the quiescence phase of novae. We have observed optical spectra and modelled those using photo-ionization code CLOUDY (version c17.00 rc1, Ferland et al., 2017). From the best-fit models, we have estimated the physical parameters e.g. temperature \& luminosity of the WD and composition \& geometry of the disk, and the type of secondaries. The results are discussed in section \ref{section4}.
The details of the observations are described in section \ref{section2} and the modelling procedure is discussed in \ref{section3} and finally, we have summarized the work in section \ref{section5}.

%%******************************* BEGINNING OF TABLE 1 *************************************************************
\begin{table*}
\centering
\caption{Log of HCT observations}
\smallskip
\centering
\begin{tabular}{l c c c c}
\hline
\hline\noalign{\smallskip}
Date						&	Julian Date	&	Object Name		&  V Magnitude	& Exposer Time	(s)		\\
\noalign{\smallskip}\hline\noalign{\smallskip}
April 28, 2015			&		2457140	& T CrB			&	10		&	360	\\
May 25, 2015			&		2457167	& RS Oph   		&	12		&	900	\\
July 23, 2015			&		2457226	& RS Oph			&	12		&	900	\\
January 31, 2016		& 		2457418	& GK Per 			&	13.5	&	1800	 \\
							& 		 			& RS Oph			&	12		&	900\\
March 17, 2016		&		2457464	& T CrB			&	10		&	360	\\
April 28, 2016			& 		2457506	& T CrB			&	10		&	360	\\
							& 					& RS Oph			&	12		&	900	\\
May 13, 2016			&		2457521	& BX Mon			&	12		&	480	\\
							& 					& T CrB			&	10		&	360	\\
June 9, 2016			&		2457548	& BX Mon			&	12		&	480	\\
September 5, 2016	& 		2457636	& T CrB			&	10		&	360  \\
							& 					& GK Per			&	13.5	&	1800  \\
November 1, 2016	&		2457693	& BX Mon			&	12		&	480	  \\
December 31, 2016	&		2457753	& BX Mon			&	12		&	480  \\
							& 					& GK Per			&	13.5	&	1800\\
							& 					& T CrB			&	10		&	360	\\
\noalign{\smallskip}\hline
\hline
\label{Observational-log}
\end{tabular}
\end{table*}
%********************************** END OF TABLE 1 *****************************

\section{Dataset}
\label{section2}
For the present study, we have taken observations using $2 m$ Himalayan Chandra Telescope (HCT), 
situated in Hanle, Ladakh, operated by Indian Institute of Astrophysics (IIA). 
 The observing nights are achieved through the submission of observation proposal of ``Optical and NIR observations of 
 Classical, Recurrent and Symbiotic Novae". During this time period, we have 
 observed and taken spectra of a handful of old novae, new novae and novae-like variables in optical (3400 - 9200 $\AA$) and near-infrared (1.00 - 2.50 $\mu m$) region.
Out of these, we present here, results of optical studies of T CrB, GK Per, RS Oph \& BX Mon. The log of observations is presented in Table \ref{Observational-log}. \\
\\
The HCT optical spectra were obtained using the Himalayan Faint Object Spectrograph Camera (HFOSC) through Grism \#7 (Resolution, R $\sim$ 1500) and Grism \#8 (R $\sim$ 2200) covering the wavelength ranges of 3400 - 8000 $\AA$ and 5200 - 9200 $\AA$ respectively ({\it https://www.iiap.res.in//iao/hfosc.html})  (Prabhu 2014).
 All spectra are reduced following the standard manner using the various tasks under IRAF (Image Reduction and Analysis Facility)
package ({\it https://iraf.noao.edu}). The spectra are bias subtracted, flat-field corrected, and cosmic rays are removed. After that, the spectra are extracted
using the optimal extraction method in the APALL task in IRAF. The wavelength calibration is done using the arc lamp observations, Iron-Argon (FeAr) for
Grism \#7 and Iron-Neon (FeNe) for Grism \#8,  and the flux calibration is done using the sensitivity function, calculated
from the spectroscopic data of standard stars Feige 34 (O type subdwarf) and Feige 110 (OB type subdwarf), observed on same night. Finally, the extracted one-dimensional spectra for Grism 7
and Grism 8 are combined using the task {\it scombine} by matching the common region between them. All the spectra are then dereddened for the interstellar extinction using the corresponding colour excess values of the systems (see Table \ref{object-list}).
All the spectra are normalized with respect to H$\beta$ line for convenience. \\
\\
In addition to these objects, we have also modelled published spectra of two other novae, V3890 Sgr and V745 Sco (Anupama \& Mikolajewska 1999). These data (CCD spectra) were obtained from the Vainu Bappu Observatory (VBO), IIA, using the $1.02 m$ and $2.3 m$ telescopes.  A brief description of these six objects is presented in Table \ref{object-list}.

%----------------------------------------------TABLE 2 ------------------------------------------
\begin{table*}
\centering
\caption{Brief description of the objects}
\smallskip
\begin{threeparttable}
\centering
\begin{tabular}{c c c c c cccccccccccccccc}
\hline
\hline\noalign{\smallskip}
Object &	Type &	Mass $^{ref.}$ & Spectra Taken	&  E(B - V)  $^{ref.}$				&	Year of Outburst & Reference \\
\noalign{\smallskip}\hline
\noalign{\smallskip}
T CrB 		& Recurrent nova				& 1.37 $M_\odot$ \tnote{a} 	     & March 17, 2016	&	0.15 \tnote{1}	&1866, 1946 			& 	Campbell 1948\\
GK Per		& Classical nova\tnote{*}	& 0.87 $M_\odot$ \tnote{b} 		 & January 31, 2016&	0.30  \tnote{2}	&1901					& Williams 1901\\
RS Oph 	& Recurrent nova				&  1.37 $M_\odot$ \tnote{a,c} 	 & January 31, 2016	&	0.75  \tnote{3}	&1898, 1907, 1933,	& Starrfield et al. 1985; \\
				& 										&											 			 &							&							&1945, 1958, 1967, 	& Kato \& Hachisu 2012\\
				&										&											 			 & 						&							&1985, 2006			&\\
BX Mon	 	& Symbiotic star					&  0.55 $M_\odot$ \tnote{d} 	 	 & May 13, 2016		&	0.20  \tnote{4}	&-							& Mayall 1940\\
V3890 Sgr	& Recurrent nova				& 1.35 $M_\odot$ \tnote{a} 		 & March 21, 1998	&	-						&1962, 1990, 2019			& Dinerstein 1973;\\
				&										&														 &							&							&						& Kilmartin \& Gilmore 1990;\\
				&										&														 &							&							&						& Strader et al. 2019\\
V745 Sco	&	Recurrent nova				& 1.35 $M_\odot$ \tnote{a} 		 & March 20, 1998	&	-						&1937, 1989, 2014	& Duerbeck 1989; Waagen 2014\\
\noalign{\smallskip}\hline
\hline
\label{object-list}
\end{tabular}
\begin{tablenotes}
\item[*] Classical nova GK Per experienced a nova ouburst on 1901 February 21 and after a long period of irregular fluctuations, it began to act like a dwarf nova (DN) since 1948. These DN outbursts generally last upto 2 months with an amplitude variation $\sim$ 1 - 3 mag (Sabbadin \& Bianchini, 1983).\\
\item[] $^1$ Selvelli et al. 1992; ~	$^2$ Bianchini \& Sabbadin 1985; ~$^3$ Mondal et al. 2018; ~$^4$ Viotti et al. 1986.
\item[] $^{a}$ Hachisu \& Kato 2001; ~$^{b}$ Wada et al. 2018; ~$^c$ Kato 2002; ~	$^d$ Dumm et al. 1998.
\end{tablenotes}
\end{threeparttable}
\end{table*}
%----------------- TABLE 2---------------------------------------------

%********************************* FIGURE 1 *********************************************
\begin{figure*}
\centering
\includegraphics[width=7 in, height =5.7 in, clip]{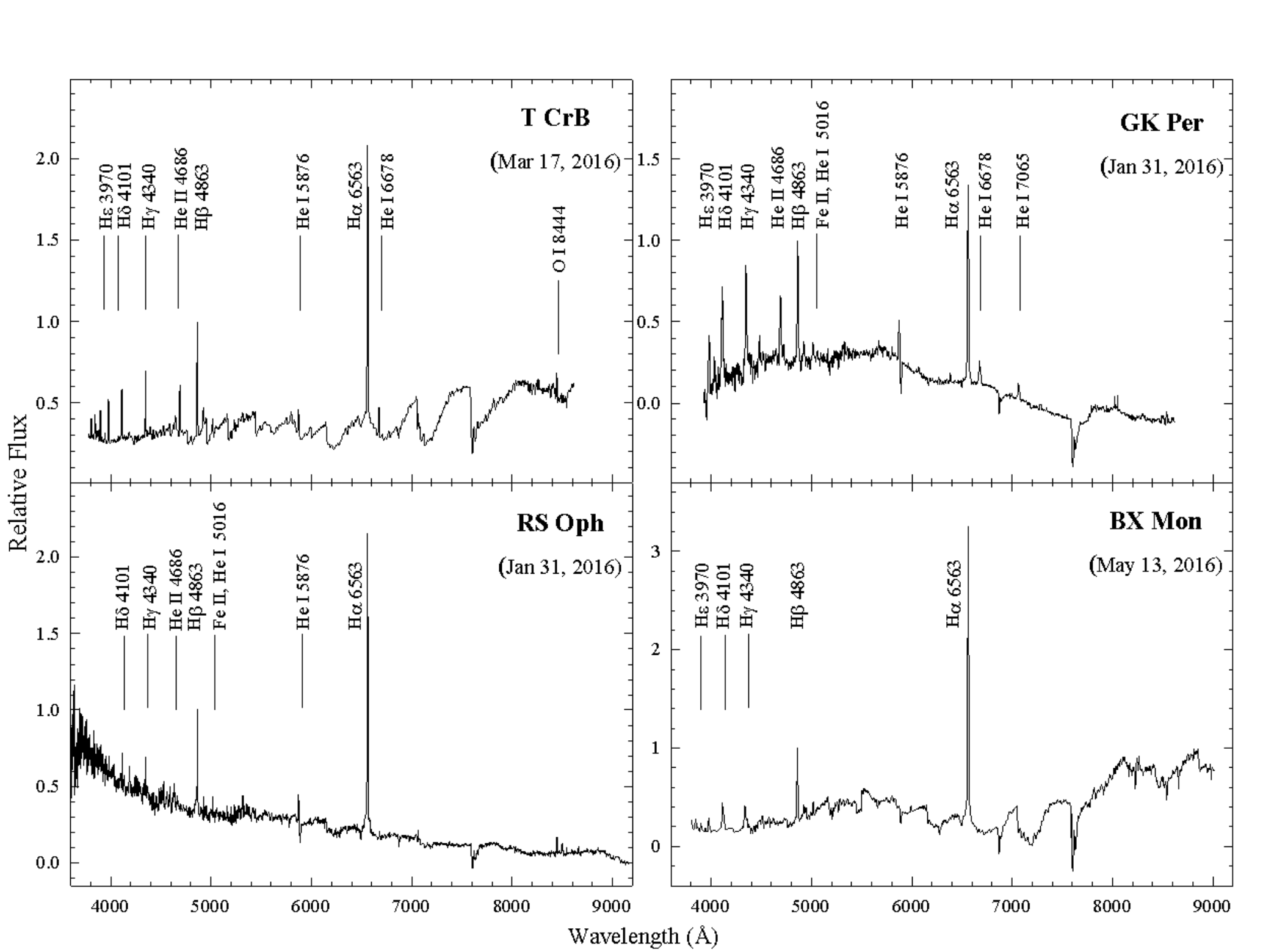}
\caption{Quiscence phase spectra of novae T CrB, GK Per, RS Oph, and symbiotic star BX Mon. The dates of observations have been mentiond in parenthesis. The spectra are normalised w.r.t. H$\beta$ line. Some of the prominent emission lines are marked in the figure. For more details see section \ref{subsection4.1}.}
\label{Qui-obs}
\end{figure*}
%********************************* FIGURE 1*********************************************

\section{Modelling procedure}
\label{section3}
The emission line spectra are modelled using CLOUDY photoionization code (version c17.00 rc1, Ferland et al., 2017). Previously, CLOUDY was used by various researchers to determine the elemental abundances and other physical characteristics of novae during the outburst, such as RS Oph (Das $\&$ Mondal 2015, Mondal et al. 2018), V1065 Cen (Helton et al. 2010), 
V1974Cyg (Vanlandingham et al. 2005), QU Vul (Schwarz  2002), as well as during quiescence (He et al. 2013).
In a similar way, we have modelled the novae atmosphere and generated synthetic spectra to estimate the physical parameters. A good understanding of the novae structure in quiescence phase i.e. the temperature and luminosity of the central source, density structure and elemental abundance of the accretion disk, and the type of secondary can be obtained using the photoionization method. 
Photoionization code CLOUDY uses a set of input parameters to compute the ionization, thermal, and chemical state of a non-equilibrium gas cloud, illuminated by a central source and predicts the resulting spectra. The input parameters include the brightness or luminosity and temperature of the radiation field striking the cloud, hydrogen density, filling factor, and composition of the gas in the accretion disk, defined by the inner ($R_{in}$) and outer ($R_{out}$) radii. The hydrogen density $n_H(r)$, and filling factor $f(r)$ of the disk may vary with the radius, $r$, as given by the following relations,
\\
\begin{equation}
n_H(r) = n_H(r_0) (r/r_0)^{\alpha} cm^{-3} \,\,\,\,\, \,\,\&\,\,\,\,\,\, f(r) = f(r_0) (r/r_0)^{\beta},
\end{equation}
\\
where $r_{0}$ is the inner radius, $\alpha$ and $\beta$ are exponents of power laws.
We have chosen the filling factor = 0.1 and filling factor power-law exponent, ($\beta$) = 0, which is the typical value used in other CLOUDY studies (e.g., Schwarz 2002, Vanlandingham et al. 2005, Helton et al. 2010). The parameter values are estimated from the best-fit model spectra and the goodness of fit is determined by the $\chi^{2}$ and reduced $\chi^{2}$ ($\chi^{2}_{red}$) of the model, given by,

\begin{equation}
\chi ^{2} = \sum\limits_{i=1}^n (M_{i} - O_{i})^{2}/ \sigma_{i}^{2}, \,\,\,\,\&\,\,\,\, \chi^{2}_{red} = \chi^{2}/\nu,
\end{equation}

where $n$ = number of observed lines, $n_{p}$ = number of free parameters, $\nu$ = degree of freedom = $n - n_{p}$, $M_{i}$ = the modelled ratio of line flux to hydrogen line flux, $O_{i}$ = measured flux ratio, and $\sigma_{i}$ = error in the observed flux ratio. Depending upon the intensity of a spectral line, possibility of blending with adjacent lines, and error in the measuring line fluxes, the error is considered in the range of $10 - 30\%$.  Value of $\chi^{2} \sim \nu$ and value of $\chi^{2}_{red}$ should be low (typically, $ 1 < \chi^{2}_{red} < 2 $) for a good fit.

\section{Results}
\label{section4}
\subsection{Description of Observed data}
\label{subsection4.1}
The observed spectra are presented in Fig. \ref{Qui-obs}. Dates of observations are mentioned on the top right corner in each panel.
The observed spectra show prominent emission features generated from the illuminated accretion disk surrounding the WD, 
which is also the source of the strong blue continuum. Generally, the strong, broad emission lines of hydrogen, helium, oxygen, and iron are seen in the spectra. 
The emission lines seen in these observed quiescence spectra are broad due to expanded disks (Anupama \& Prabhu 1989; Anupama 2008). Strong optical Fe II emission lines come from an illuminated (photoionized) high density inner region of the disk. As the disk accumulates more matter, the lines weaken (Bode \& Evans 2008).
The spectra also show prominent absorption features in the continuum due to the cool secondaries and they reveal the properties of the secondaries. \\
\\
The strong emission features observed in the spectrum of T CrB (observed on March 17, 2016) are due to H I (Balmer emissions) lines along with He I 5016, 5876, 6678 $\AA$, He II 4686 $\AA$, Fe II 5016 $\AA$ and O I 8444 $\AA$ lines. 
The absence of higher ionization lines can be accounted for by the absorption and softening by reradiation of all direct photons from the accretion disk. Presence of N III 4640 $\AA$\ and Ca II 3933 $\AA$ along with He II 4686 $\AA$ indicate the super active phase that took place in October 2015. The strong TiO absorption features at 6180 and 7100 $\AA$ are due to the secondary.\\
\\
The spectrum of GK Per was observed on January 31, 2016, about ten months after its recent dwarf nova (DN) outburst on March 2015 (Wilber et al. 2015), which lasted for 2 months. 
Presence of strong He I 5876, 6678 \& 7065 $\AA$, and He II 4686 $\AA$ lines and Fe II features along with strong Balmer emission lines indicate towards this active phase and the bluer continuum suggests the higher temperature of the accretion disk. TiO absorption features due to the cool component are relatively weak.\\
\\
The quiescence-phase spectrum of RS Oph taken on January 31, 2016, shows prominent emission lines of H (Balmer emissions) and He I 4471, 5016, 5876, 6678 \& 7065 $\AA$, together with strong TiO absorption features at 5448 and 6180 $\AA$ contributed by the secondary.
We observed that in comparison with the spectrum observed on April 26, 2007 (Mondal et al. 2018), Fe II lines are weaker
 which indicate that the accretion disk has been formed.
This is in line with the fact that the mass accretion rate is high in the case of RS Oph (Yaron et al. 2005).\\
\\
The spectrum of BX Mon (observed on May 13, 2016) is shown in the fourth panel of Fig. \ref{Qui-obs}. The symbiotic star BX Mon was discovered by Mayall (1940) and its classification is based on the combination spectra in the optical region showing
simultaneously very high excitation emission lines and low temperature absorption features (Kenyon 1986). 
No nova-like eruption has ever been recorded for this system. However, the eclipsing binary system shows variability with a periodicity of 1401 days and an amplitude of $\sim$ 3 magnitudes (Dumm et al. 1998).
The observed spectrum is dominated by the strong TiO absorption features at 5448 and 7100 $\AA$ contributed by the cool component. Balmer emission lines are seen superimposed on the spectrum of the cool red continuum. No other strong emission lines are seen in the observed spectrum. This is possibly due to the low temperature of the disk. 

 %--------------------------- TABLE 3 ------------------------------------------------------------
%
\begin{table*}
%\tiny
\centering
\caption{Observed and best-fit CLOUDY model line fluxes relative to H$\beta$}.
\smallskip
\centering
\begin{tabular}{c c|c c c |c c c|c c cccccccc}
\hline
\hline\noalign{\smallskip}
$\lambda$ &	      &     			&  T CrB   &            &       &	 GK Per  &	 	       &      &  RS Oph  &  \\
 ($\AA$)  & Line Id & Obs. &  Mod.    & $\chi^{2}$ & Obs.  &  Mod.     & $\chi^{2}$ & Obs.  &  Mod.     & $\chi^{2}$  \\
\noalign{\smallskip}\hline\noalign{\smallskip}
3889    & H$\zeta$, He I  		& 0.21 & 0.16	& 0.25	& ...	  &...		&...		& ...	  &...		&...		\\
3970    & H$\epsilon$, 			& 0.34 & 0.27	& 0.49	& 0.41 & 0.33	& 0.64	& ...	  &...		&...		\\
4101    & H$\delta$   				& 0.52 & 0.43	& 0.81	& 0.63 & 0.53	& 1.00	& 0.37 & 0.40	&0.09\\
4340    & H$\gamma$       		& 0.61 & 0.55	& 0.36	& 0.80 & 0.72	& 0.64	& 0.43 & 0.52	&0.81\\
4471    & He I 	          			& ...	  &...		&...		& ...	  &...		&...		& 0.08 & 0.11	& 0.09	\\
4686    & He II 						& 0.37 & 0.40	& 0.09	& 0.62 & 0.57	& 0.25	& ...	  &...		&...		\\
4863	   & H$\beta$					& 1.00 & 1.00	& 0.00  	& 1.00 & 1.00	& 0.00	& 1.00 & 1.00	& 0.00\\
5016    & Fe II, He I				& 0.45 & 0.53	& 0.64	& ...	  &...		&...		& 0.12 & 0.20	& 0.64		\\
5876    & He I        				& 0.11 & 0.20	& 0.81	& 0.31 & 0.30	& 0.01	& 0.20 & 0.31	& 1.00		\\
6563    & H$\alpha$   				& 3.06 & 3.22	& 2.56	& 1.92 & 2.57	& 4.69	& 3.30 & 3.43	& 1.69\\
6678    & He I        				& 0.13 & 0.18	& 0.25	& 0.09 & 0.13	& 0.16	& 0.07 & 0.12	& 0.25		\\
7065    & He I		  					& ...	  &...		&...		& 0.07 & 0.10	& 0.09	& 0.08 & 0.14 & 0.36\\
8444    & O I							& 0.09 & 0.16	& 0.49	& ...	  &...		&...		&...&...&...\\
\noalign{\smallskip}\hline\noalign{\smallskip}
Total &									&...	   &...	&6.75	& ...  &... &7.48	&...&...	&4.93\\
\noalign{\smallskip}\hline
%&&&&&&&&&&\\
\hline\noalign{\smallskip}
$\lambda$ &	       					&      &  BX Mon  &  &      & V3890 Sgr & &      &  V745 Sco &  \\
 ($\AA$)  & Line Id 					& Obs. &  Mod.    & $\chi^{2}$ & Obs.  &  Mod.     & $\chi^{2}$ & Obs.  &  Mod.     & $\chi^{2}$  \\
\noalign{\smallskip}\hline\noalign{\smallskip}
3970    & H$\epsilon$, 			& 0.21 & 0.27	& 0.36	& ...	  &...		&...		& ...	  &...		&...		\\
4101    & H$\delta$   				& 0.39 & 0.35	& 0.16	& ...	  &...		&...		& ...	  &...		&...		\\
4340    & H$\gamma$       		& 0.32 & 0.43	& 1.21	& 0.25 & 0.30	& 0.25	& ...	  &...		&...		\\
4471    & He I 	          			& ...	  &...		&...		& ...	  &...		&...		& ...	  &...		&...		\\
4686    & He II 						& ...	  &...		&...		& ...	  &...		&...		& ...	  &...		&...		\\
4863	   & H$\beta$					& 1.00 & 1.00	& 0.00  	& 1.00 & 1.00	& 0.00	& 1.00 & 1.00	& 0.00\\
5016    &  Fe II, He I				& ...	  &...		&...		& 0.38 & 0.41	& 0.09	& 0.17 & 0.22	& 0.25\\
5169	  &  Fe II						& ...	  &...		&...		& 0.49 & 0.44	& 0.25 	& 0.18 & 0.21 & 0.09\\
5317	  &  Fe II						& ...	  &...		&...		& 0.18 & 0.21	& 0.09	& ...	  &...		&...	\\	
5876    & He I        				& ...	  &...		&...		& 0.40 & 0.44	& 0.16	& 0.61 & 0.50	& 1.21 		\\
6563    & H$\alpha$   				& 3.95 & 4.08	& 1.69	& 8.25 & 7.36	& 8.80 	& 6.83 & 6.72	& 1.21\\
6678    & He I        				& ...	  &...		&...		& 0.18 & 0.20	& 0.04	& 0.24 & 0.30	& 0.36		\\
7065    & He I		  					& ...	  &...		&...		& 0.15 & 0.18	& 0.09	& 0.18 & 0.22 & 0.16\\
\noalign{\smallskip}\hline\noalign{\smallskip}
Total &							&...	   &...	&3.42	& ...  &... &9.77	&...&...	&3.28\\
\noalign{\smallskip}\hline
\hline
\label{chisq-table-2}
\end{tabular}
\end{table*}
%%%%%------------------------------------------TABLE 3-----------------------------

\subsection{Modelling of spectra}
\label{subsection4.2}
\begin{figure*}
\centering
\includegraphics[width=7 in, height =8 in, clip]{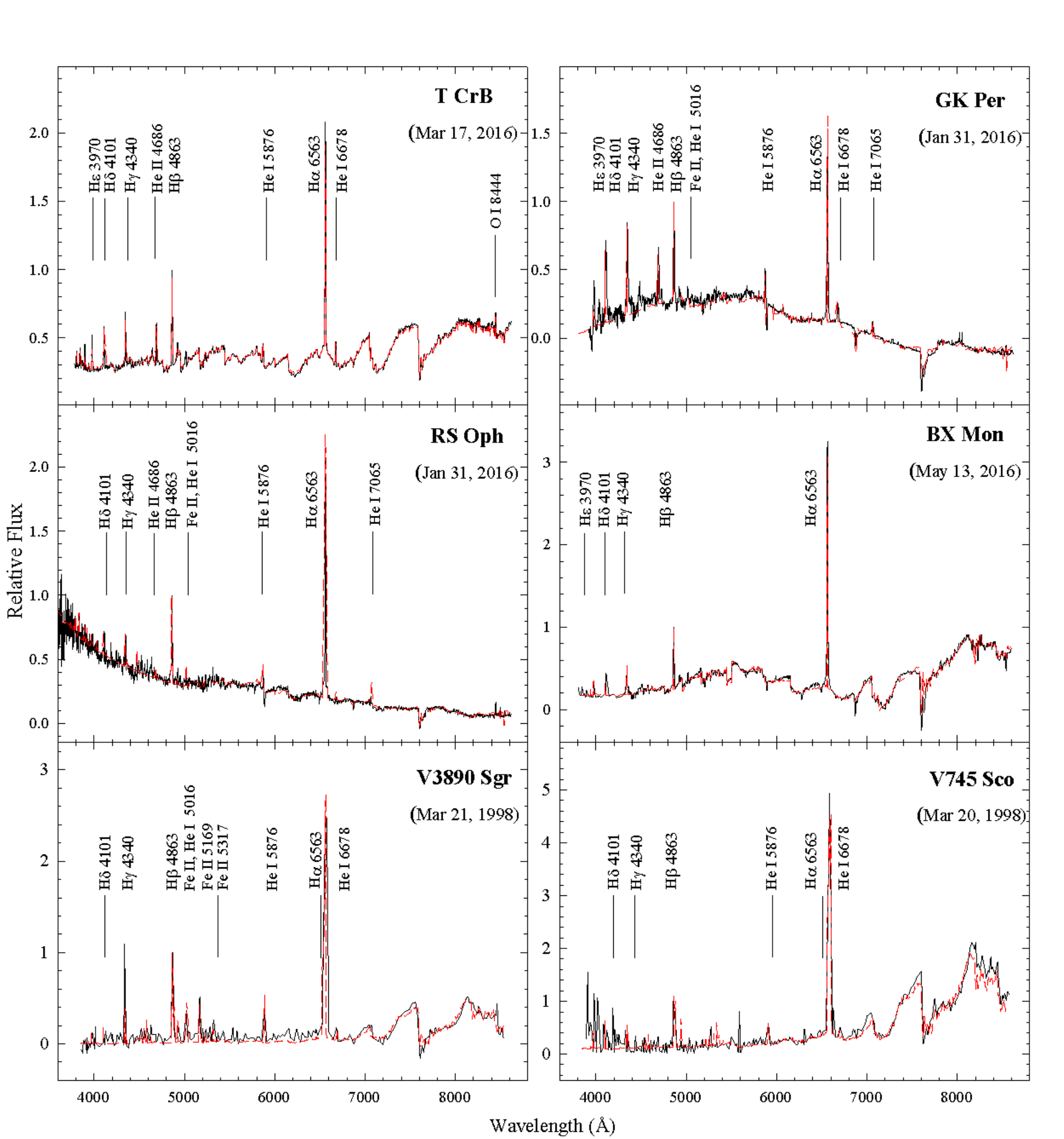}
\caption{Observed (black solid) and model generated spectra (red dashed line) of quiscence phase of novae T CrB, GK Per, RS Oph, V3890 Sgr, V745 Sco, and symbiotic star BX Mon. Dates of observation for each system have been mentioned in  parenthesis. All the spectra are normalized w.r.t. H$\beta$ line. Some of the prominent emission lines are marked in the figure. For more details see section \ref{subsection4.2}.}
\label{Qui-mod}
\end{figure*}

For modelling of the spectra, we have considered a three component model consisting of the radiation from the central WD, photoionization emission from the accretion disk formed around the WD and the cool continuum coming from giant secondary. \\
\\
To reduce the number of free parameters, we have kept the $R_{in}$ and $R_{out}$ values fixed for the objects as the number of emission lines are less in the quiescence spectra. The values of $R_{in}$ and $R_{out}$ have been calculated in the following way.
Hamada \& Salpeter (1961) estimated radius of a 1 $M_\odot$ WD as $10^{8.9} cm$. Paczynski (1977) showed that outer radius of the accretion disk can reach up to $3 \times 10^{11} cm$  for a close binary system consisting of a 1 $M_\odot$ WD, 0.5 $M_\odot$ secondary and having orbital period of 3 - 10 $hr$. Combining these resutls, 
for simplicity of our calculations, we consider $R_{in}$ and $R_{out}$ of the accretion disk around a 1 $M_\odot$ WD  as $10^{8.9} cm$ and $10^{11} cm $ respectively.
Using these values and the relation between radius ($R_{WD}$) and mass ($M_{WD}$) of a WD, $R_{WD} \propto M_{WD}^{-1/3}$ (considering non-relativistic degeneracy), we have estimated the values of $R_{in}$ and $R_{out}$ for different systems.\\
\\
Also, the semi-height of the cylindrical disk has been kept fixed as $10^8 cm$ for the same reason. 
A constant mass per unit volume throughout the model shell gives the value of $\alpha$ = - 2.  From the H$\alpha$/H$\beta$ line flux ratio, Hydrogen density of the disk is estimated. The rest of the elements are kept at solar abundance except for those which show strong presence of elemental lines in spectra. By matching the elemental emission lines of observed and modelled spectra, elemental abundances are determined.
The spectra for giants are taken from the archival data of $1.5 m$ SMARTS (Small \& Moderate Aperture Research Telescope System) telescope in Cerro Tololo Inter-American Observatory (CTIO), Cerro Tololo, Chile (Riedel et al. 2007) and $4.5 m$ Multiple Mirror Telescope (MMT Observatory) on Mount Hopkins, Arizona, USA (Henry et al. 1994).\\
\\
These three components (black-body radiation of the WD, model generated emission line spectra and absorption spectra of giants) are then added up in various ratios to get the final best-fit model spectra using the $\chi^{2}$ minimization technique. Both the observed and model generated spectra are normalized with respect to H$\beta$ line. Observed (black solid) and modelled spectra (red dashed) are shown in Fig. \ref{Qui-mod}. The relative fluxes of the best-fit model predicted lines, observed lines and corresponding $\chi^2$ values are presented in Table 
\ref{chisq-table-2}; and the best-fit model parameters are presented in Table \ref{Quiescence-Parameters}. From the best-fit model spectra, contribution of the components are also determined. In the following subsections, details of the parameters for individual nova are discussed.\\
\\
%********************************* FIGURE 2 *********************************************
%********************************* FIGURE 2*********************************************
%%%%%%%%%%%%%%%%%%%%%%%%%%%%%%%%%%%%%
{\bf T Coronae Borealis (T CrB):} 
For T CrB, we have considered a high mass hot WD (1.37 $M_\odot$, $T = 10^{4.1} K$) surrounded by a cylindrical accretion disk ($R_{in} = 10^{8.85} cm$, $R_{out} = 10^{10.95} cm$) of thick H-rich material ($n_H = 10^{11.2} cm^{-3}$).
Presence of higher ionization lines (e.g. N III, He II, Ca II) indicate the presence of some hot centres in the accretion disk. Thus, for the modelling, we considered two shells with different temperatures ($T = 10^5 K$, $L = 10^{32} erg/s$ for higher ionization lines and $T = 10^{4.1} K$, $L = 10^{34} erg/s$ for lower ionization lines) and added them to match the emission features.
The absorption features seen in the observed spectrum are due the secondary. 
Previously, Kleinmann \& Hall (1986) suggested a spectral type of M2-M5 III and M\"{u}rset \& Schmid (1999) suggested a spectral type of M4.5 III for the secondary of T CrB. Anupama \& Mikolajewska (1999) determined a spectral type of M3 - 4 III. We added the spectrum of M4 type giant with the modelled spectrum and found that it fits well with the absorption features. As the properties of M4 type giant ($T \sim 3500 K$, $L \sim 7 L_\odot$) does not contribute any emission lines in the spectra, all the emission features for the T CrB spectrum are generated from the accretion disk. So only the model spectrum generated from the disk added with the black-body radiation from the WD was sufficient to produce the emission lines.
By fitting the emission lines of different elements with varying abundances, we estimate the elemental abundances using $\chi^2$ values. Elemental abundances of He, Fe and O are estimated using the prominent emission features. The strong Fe II 5016 $\AA$ line is blended with He I. But as He abundance is estimated from other strong He I lines, we consider the blended Fe II line to estimate the Fe II abundance.\\
\\
{\bf GK Persei (GK Per):}
We considered a 0.87 $M_\odot$ WD for modelling GK Per spectra and calculated $R_{in}$ and $R_{out}$ as $10^{8.92} cm$ and $10^{11.02} cm$.
To generate higher ionization lines (e.g. He II), we had to adopt a higher effective temperature for the source. The best-fit model reveals $T = 10^{4.5} K$, $L = 10^{34.5} erg/s$ and  $n_H = 10^{11.5} cm^{-3}$. The high temperature and luminosity indicate towards the active phase of the dwarf nova outburst in March 2015.
From the He I and He II lines, abundance of helium is estimated as 0.8 with respect to solar abundances, which is subsolar. Abundances of other elements could not be estimated due to the absence of prominent lines. Absence of Fe II features can be accounted for a thick disk as the Fe II lines are softened by the optically thick, cooler outer region of the disk. 
Earlier, Gallagher \& Oinas (1974) determined the type of secondary as K2 V - IV and Morales-Reuda et al. (2002) found the donor star type as K1 IV.
In this case, we observed that the spectrum of K5 V giant fits well with the absorption features seen in the observed spectrum. So, a spectrum of K5 V giant was added with the CLOUDY modelled spectrum to get the final spectrum.\\
\\
{\bf RS Ophiuchi (RS Oph):}
RS Oph system is composed of a massive WD ($\sim$ 1.37 $M_\odot$) primary accompanied by a red giant secondary. The radii of accretion disk are calculated as 
$R_{in} = 10^{8.85} cm$ and $R_{out} = 10^{10.95} cm$.
From the best-fit model, we find that the hot and luminous WD ($T = 10^{4.2} K$ and $L = 10^{32.0} erg/s$) is surrounded by a dense accretion disk of H-rich material ($n_H = 10^{11.1} cm^{-3}$). A spectral type of M2 III fits the TiO absorption features of the secondary well. These two spectra are added with the black-body radiation continuum from the WD ($T = 10^{4.2} K$) to get the final model generated spectrum.
Helium abundance is estimated as 0.3 with respect to solar abundances, which is also subsolar. Only one prominent Fe II line is present but that too is blended, so, abundance of Fe cannot be estimated accurately. Since there were no prominent lines of other elements, the abundances of other elements could not be estimated. 
In previous work, Mondal et al. (2018) used CLOUDY code to model the quiescence phase spectrum of RS Oph taken on April 26, 2007 and estimated the physical parameters of the system and also estimated spectral class of the secondary as M2 III type which matches with previous results by Worters et al. (2007) and Anupama \& Mikolajewska (1999). \\
 \\
{\bf BX Monocerotis (BX Mon):}
Symbiotic star BX Mon shows similar observational behaviour as the quiescence phase nova. 
For the modelling of the spectrum of BX Mon, we have considered a similar approach of three component model consisting of WD, accretion disk and the giant and added the spectra to match with the observed one. We have considered mass of the primary as 0.55 $M_\odot$ and obtained $R_{in} = 10^{8.98} cm$ and $R_{out} = 10^{11.09} cm$ for accretion disk.
From the best-fit model, we got the hot WD ($T = 10^{4.1} K$, $L = 10^{31.5} erg/s$) is surrounded by thick H-rich ($n_H = 10^{11.4} cm^{-3}$) cylindrical accretion disk. Elemental abundance cannot be estimated as the prominent emission lines of other elements except Balmer lines are absent.
Previously M\"{u}rset \& Schmid (1999) determined a spectral type of the secondary as M4 - 5.5 III.
We added a spectrum of M4.5 III type with the modelled spectrum and found it matches well with the absorption features seen in the observed spectrum. Thus, it confirms that the secondary type for BX Mon is M4.5 III.\\
\\
%%******************************* BEGINNING OF TABLE 4  *************************************************************
\begin{table*}
\centering
\caption{Best-fit CLOUDY Model parameters}
\smallskip
\begin{threeparttable}
\centering
\begin{tabular}{l c c c c c c c}
\hline
\hline\noalign{\smallskip}
 Parameters                   				&	 T CrB 		&	GK Per & RS Oph	&BX Mon & V3890 Sgr & V745 Sco\\
\noalign{\smallskip}\hline\noalign{\smallskip}
Log($T$) ($K$)             			&   4.1/5.0 	& 4.5			& 4.2 		& 4.1			& 4.05 		& 4.02 \\
Log($L$) ($erg/s$)			&   32.0/34.0  & 34.5		& 32.0		& 31.5 		& 29.3		& 29.1 \\
Log($n_H$) ($cm^{-3}$) 		  		&	  11.2			& 11.5		& 11.1		& 11.4		& 11.2		& 11.1\\
$\alpha$\tnote{a}      						&   -2 			&  -2			&  -2			&  -2			&  -2			&  -2\\
Log($R_{in}$) ($cm$)		  				&    8.85		&	8.92		& 8.85		& 8.98		& 8.86		& 8.86\\
Log($R_{out}$) ($cm$)          			&   10.95		& 11.02		& 10.95		& 11.09		& 10.96		& 10.96\\
Log(semi-height) ($cm$)						&    8   			&	8			& 8			& 8			& 8			& 8\\
Filling factor	              					&	 0.1     		&  0.1		&  0.1		&  0.1		&  0.1		&  0.1\\
$\beta$\tnote{b}		      				&	 0.0    		&  0.0		&  0.0		&  0.0		&  0.0		&  0.0\\
He/He$_\odot$\tnote{c}        			&  0.5(6) 		&	0.8(4)	& 0.3(5)		& --			& 1.5(4)		& 1.1(4)\\
Fe/Fe$_\odot$\tnote{c}        			&  8.0(1)		&	-- 			& 1.8	(1)	& --			& 10.0(3)	& 10.0(2)\\
O/O$_\odot$\tnote{c}        			&  3.0(1)		&	-- 			& --			&	-- 			& --			&	-- 	\\
Number of observed lines ($n$)  			&   11      		& 9			& 	9			& 5			& 9			& 7\\
Number of free parameters ($n_{p}$)&  6  			&	4			& 5			& 3			& 5			& 5\\
Degrees of freedom ($\nu$)		  		&   5  			&	5			& 4			& 2  			& 4			& 2\\
Total $\chi^{2}$ 	              			& 	6.75			& 7.48		& 4.93		& 3.42		& 9.77		& 3.28\\
$\chi^{2}_{red}$                  			& 	 1.35			& 1.49		& 1.23		& 1.71		& 2.44		& 1.64\\
Type of secondary							&  M4 III		& K5 V		& M2 III	& M4.5 III & M5 III 	& M5.5 III\\
Contribution of secondary					& 40\%			& 30\%		& 35\%		& 55\%		& 75\%		& 70\%\\
\noalign{\smallskip}\hline
\hline
\label{Quiescence-Parameters}
\end{tabular}
\begin{tablenotes}
\item[a] Radial dependence of the density $r^{\alpha}$
\item[b] Radial dependence of filling factor $r^{\beta}$
\item[c] Abundances are given in logarithmic scale, relative to hydrogen. All other elements which are not listed
in the table were set to their solar values. The number in the parentheses represents number of lines used in
determining each abundance.
\end{tablenotes}
\end{threeparttable}
\end{table*}
%********************************** END OF TABLE 4 **************************
{\bf V3890 Sagittarii (V3890 Sgr):}
We used the published optical spectroscopic data of V3890 Sgr observed at VBO on March 21, 1998 (Anupama \& Mikelojewska, 1999).
The spectrum shows prominent Balmer emission lines along with strong Fe II lines, indicating the accretion disk is not fully formed. Absence of higher ionization lines marks the quiescence phase of the system. The absorption features are due to the giant secondary. Using the previously calculated value of the mass of WD primary as 1.35 $M_\odot$, we find $R_{in} = 10^{8.86} cm$ and $R_{out} = 10^{10.96} cm$ for accretion disk.
From the best-fit CLOUDY model, we concluded that the system consists of an ionizing source of low temperature ($T = 10^{4.05} K$) and luminosity ($10^{29.3} ergs/s$) surrounded by a cylindrical accretion disk of H-rich material ($n_H = 10^{11.2} cm^{-3}$). Using the prominent He I (5016, 5876, 6678 $\&$ 7065 $\AA$) and Fe II (5016, 5159 $\&$ 5317 $\AA$) lines, elemental abundances of helium and iron are estimated as 1.5 and 10 respectively with respect to solar abundances. 
Also, a spectral type M5 III added to the model generated spectrum to match the absorption features and we found that the spectrum fits well. Anupama \& Mikelojewska (1999) determined the type of the secondary as M5 III. Harrison et al. (1993) also indicated a spectral type M5 III based on the infrared photometric data during 1990 outburst. Our result matches with the previous results. In general, we use IRAF software to measure the flux of emission lines, but in case of the observed spectrum of V3890 Sgr, H$\alpha$ line is truncated. So, we have used line fluxes measured by Anupama \& Mikelojewska (1999).
Probably due to the difference in method for the measurement and presence of fewer emission lines, we obtained a higher $\chi^2_{red}$ value.\\
\\
{\bf V745 Scorpii (V745 Sco):} 
To study the quiescence phase spectrum of V745 Sco, we have modelled the published optical spectroscopic data taken on March 20, 1998 (Anupama \& Mikelojewska, 1999) using CLOUDY photoionization code. The spectrum shows prominent Balmer emission lines although the blue region is not well resolved. Prominent Fe II and He I lines can also be easily identified. The strong TiO absorption features are due to the secondary companion. 
The radii for accretion disk have been calculated as $R_{in} = 10^{8.86} cm$ and $R_{out} = 10^{10.96} cm$ for a WD mass of 1.35 $M_\odot$. The best-fit model results a hot central source with relatively lower temperature ($T = 10^{4.02} K$) and luminosity ($L = 10^{29.1} ergs/s$), is surrounded by thick H-rich ($n_H$ = $10^{11.1} cm^{-3}$) cylindrical accretion disk. From the prominent He I 5016, 5876, 6678 \& 7065 $\AA$ lines and Fe II 5016 \& 5169 $\AA$ lines, elemental abundances of helium and iron are estimated as 1.1 and 10 respectively with respect to solar abundances. 
Williams et al. (1991) suggested the red giant companion of V745 Sco as M6 - 8 III type, Duerbeck (1989) classified it as M6 III and Harrison et al. (1993), based on the infrared spectra inferred a spectral type of M4 III. Based on the TiO and VO indices of the optical spectra, Anupama \& Mikelojewska (1999) suggested a spectral type of M4 - 5 III, while the Na I index indicated towards a M6 III type. From our study, we have observed that the spectrum of a M5.5 III type giant fits the absorption features well. So, a spectrum of M5.5 III type giant is added to get the final spectrum.

\section{Summary \& Discussions}
\label{section5}
We have studied spectroscopic behaviour of five novae (T CrB, GK Per, RS Oph, V3890 Sgr, V745 Sco) and one symbiotic star (BX Mon) in the optical region. 
 We have used spectra observed at $2m$ HCT and published data. We have modelled these quiescence phase spectra using photoionization code CLOUDY. 
 We have generated spectra emitted from the accretion disk; and added these model generated spectrum with black-body continuum corresponding to the WD temperature and secondary spectrum to get the final spectra. Generally single shell model has been sufficient but in case of T CrB, we had to consider two different shells corresponding to higher and lower temperatures to match the higher (N III, Ca II) and lower (He I, Fe II, O I) ionization lines. From the best-fit models, the physical parameters, e.g. temperature of the WD; temperature, luminosity, H-density of the disk etc. related to the system have been estimated. In general the WD temperature and luminosity are in the range of $T = 10^{4.02 - 4.5} K$ ($T = 10^{5.0} K$ for higher ionization lines in T CrB) and $L = 10^{29.1 - 34.5} erg/s$; density of the disk is $n_H = 10^{11.1 - 11.5} cm^{-3}$. For GK Per, the temperature is high ($T = 10^{4.5} K$); presence of strong He II line indicates that the system might be still in its active phase. 
Weak presence of Fe II lines in the observed spectra suggests that for GK Per \& RS Oph the disk is thick, whereas, strong Fe II lines in the spectra of V3890 Sgr \& V745 Sco indicate that the disk is not fully formed. For the same reason, luminosity values for V3890 Sgr \& V745 Sco are on the lower side.\\
\\
We also have estimated abundances for few elements, e.g., He, Fe, \& O for the objects where emission lines of those elements were prominent. 
We find He abundance is sub-solar in case T CrB, GK Per and RS Oph  and super-solar in V3890 Sgr and V745 Sco. Fe abundance, estimated for T CrB, RS Oph, V3890 Sgr and V745 Sco, and oxygen abundance that has been estimated for T CrB, are found higher in these novae. Our results may be compared with few previous studies. For example, 
Mondal et al. 2018 found higher abundance of He (He/He$_\odot$ = 1.6 - 2.5) and Fe (Fe/Fe$_\odot$ = 0.5 - 3.8) in RS Oph spectra obtained during 2006 outburst, whereas in the present study we find lower values of He and Fe abundances (He/He$_\odot$ = 0.3 and Fe/Fe$_\odot$ = 1.8) in the quiescence phase. This is reasonable as the metallicity increases during outbursts due to dredging up and mixing of WD materials
(Jos\'e \& Hernanz 1998, Starrfield et al. 1998) that enrich the accreted envelope. 
In case of GK Per, our estimation of He abundance is roughly consistent with that obtained by Anupama \& Prabhu (1993), who found He abundance as near solar from its quiescence phase spectra.\\
\\
By fitting the spectra of various secondaries and matching the strength of TiO absorption bands, we estimated the type of secondaries and percentage of their contributions for these objects. For T CrB, RS Oph, BX Mon, V3890 Sgr \& V745 Sco the secondaries are in the range of M2 III - M5.5 III types and for GK Per the secondary is K5 V type. These results are in line with the fact that recurrence period of the systems with M-type giants are $\sim$ 20 - 80 yrs as the mass loss rate of M type giants $\sim$ 10$^{-5}$ - 10$^{-6}$ $M_\odot/yr$ (Wannier \& Sahai 1986, Zemko et al. 2017). For the classical nova GK Per mass accretion rate is much slower ($\sim$ 10$^{-8}$ $M_\odot/yr$) (Zemko et al. 2017) due to K type secondary.\\
\\
From the best-fit model parameters, we may get rough idea about the disk size around the novae. It is estimated that the contribution of the accretion disks in the final spectra of T CrB, GK Per and RS Oph are higher in comparison to the contribution of the secondary. This is probably due to the reason that the disk size is comparatively larger. On the other hand, in case of BX Mon, V3890 Sgr and V745 Sco, contribution of the disk is lower because of its smaller size. However, we must consider this with cautions, as our model is based on simple assumptions and does not include many important parameters, e.g., inclination angle of the disk.\\
\\
For all the novae, except V3890 Sgr, reduced $\chi^2$ values lie in between 1 \& 2 which indicates that the model generated spectra matched the observed spectra satisfactorily. For V3890 Sgr, the reduced $\chi^2$ value is relatively higher, 2.44. This is probably due to the difference in method for the measurement of H$\alpha$ line, which is the highest contributor to the $\chi^2$ value.\\
\\
We are presently working on the remaining observed data and trying to understand their properties through modelling of spectra which will be published in a future work.\\

\section{acknowledgements}
We acknowledge all the technicians and staffs of the IAO (Indian Astronomical Observatory) $\&$ CREST (Centre For Research \& Education in Science \& Technology) for their kind help and support during the observations. We also thank the anonymous referee for valuable comments and suggestions that helped in improving the results presented here.

\bibliography{bibtex}

\end{document}